\documentstyle[twocolumn,aps]{revtex}
\begin{document}
\draft
\title{\large {\bf A Giant Halo at the Neutron Drip Line \cite{AAA}}}
\author{J. Meng and P. Ring}
\address{T30 Physik-Department der Technischen Universit\"at M\"unchen,\\
D-85747 Garching, FRG}
\date{\today}
\maketitle
\begin{abstract}
Relativistic Hartree-Bogoliubov (RHB) theory in coordinate
space is used to describe the chain of even-even Zirconium
isotopes reaching from $^{116}$Zr to the drip line nucleus
$^{140}$Zr. Pairing correlations are taken into account by
a density dependent force of zero range. For neutron
numbers larger than the magic number $N=82$  a giant
neutron halo outside the core of $^{122}$Zr is observed. It
is formed by up to six neutrons.
\end{abstract}
\par
\pacs{PACS numbers : 21.10.Pc, 21.10.Gw, 21.60.-n,21.60.Jz}

\narrowtext


With perpetual improvement in the techniques of producing
radioactive beams, the study of exotic nuclei very far away
from the line of $\beta$-stability has now become feasible. 
Experiments of this kind may cast new light on
nuclear structure and novel and entirely unexpected
features may appear: Neutron rich nuclei can have a
structure very different from that of normal nuclei. They
consist of a normal core surrounded by a skin  of 
neutron matter. Close to the drip line, where the coupling
to the continuum becomes important, a neutron halo can
develop, as it has been observed in several light nuclei,
the most famous being $^{11}$Li \cite{THH.85b}. However, in
all of the halos observed so far, one has only a very small
number of neutrons, namely one or two outside of the normal
core.

In order to study the influence of correlations and
many-body effects it would be very interesting to find also
nuclei with a larger number of neutron distributed in the
halo. In this letter we report on the theoretical
prediction of a giant neutron halo for Zr-isotopes close
to the neutron drip line. It is formed by up to six neutrons
outside of the $^{122}$Zr core with the magic neutron
number $N=82$.

Recently, a fully self-consistent relativistic method has
been developed for the description of halo nuclei. It is
based on Relativistic Hartree-Bogoliubov (RHB) theory in
coordinate space \cite{MR.96}. It is able to take into
account at the same time the proper isospin dependence of
the spin-orbit term, which is very crucial for a reliable
description nuclei far away from the line of
$\beta$-stability \cite{SLH.94} and the self-consistent
treatment of pairing correlations in the presence of the
continuum \cite{DFT.84}. As it has been shown in Ref.
\cite{MR.96} the scattering of Cooper pairs into the
continuum containing low-lying resonances of small angular
momentum plays an important role for the formation a
neutron halo. By using a density dependent zero range
interaction, the  halo in $^{11}$Li has been successfully
reproduced in this self-consistent picture. Very good
agreement with recent experimental data is obtained without
any adjustment of parameters. To obtain these results, a
full solution of the RHB equations \cite{KR.91} in
coordinate space is necessary. The expansion in a harmonic 
oscillator basis \cite{GEL.96}, which is very useful 
only for nuclei close to the line of $\beta$-stability,
provides only a very poor approximation in the continuum,
even if a large number of oscillator shells is taken into
account.  The simple BCS-approximation used in Ref.~\cite{SLH.94} 
leads to a partial occupation of unbound states in the continuum 
and therefore to a gas of evaporating neutrons. Here we report
on calculations, where the same method as in Ref.~\cite{MR.96} 
has been used to investigate halos in Zr-nuclei at the neutron 
drip line.

The starting point is the Relativistic Mean Field
Lagrangian
\begin{eqnarray}
{\cal L}&=&\bar\psi\left(\rlap{/}p - g_\omega\rlap{/}\omega -
g_\rho\rlap{/}\vec\rho\vec\tau - \frac{1}{2}e(1 - \tau_3)\rlap{\,/}A -
g_\sigma\sigma - m\right)\psi
\nonumber\\
&&+\frac{1}{2}\partial_\mu\sigma\partial^\mu\sigma-U(\sigma)
-\frac{1}{4}\Omega_{\mu\nu}\Omega^{\mu\nu}+ \frac{1}{2}
m^2_\omega\omega_\mu\omega^\mu\\ 
&&-\frac{1}{4}\vec R_{\mu\nu}\vec R^{\mu\nu}+
\frac{1}{2} m^2_\rho\vec\rho_\mu\vec\rho^\mu 
-\frac{1}{4}F_{\mu\nu}F^{\mu\nu}
\nonumber
\end{eqnarray}
It describes the nucleons with the mass $m$ as Dirac
spinors $\psi$ moving in the fields of mesons: a scalar
meson ($\sigma$), an isoscalar vector meson ($\omega$), an
isovector vector meson ($\vec\rho$) and the photon $A^\mu$,
with the masses $m_\sigma$, $m_\omega$ and $m_\rho$ and the
coupling constants $g_\sigma$, $g_\omega$, $g_\rho$. The
field tensors for the vector mesons are given as
$\Omega_{\mu\nu}=\partial_\mu\omega_\nu-\partial_\nu\omega_\mu$
and by similar expressions for the $\rho$-meson and the
photon. For simplicity we neglect these fields in the following, 
however, they are fully taken into account in the
calculations. For a realistic description of nuclear
properties a nonlinear self-coupling $U(\sigma) =
\frac{1}{2} m^2_\sigma \sigma^2_{}
+\frac{1}{3}g_2\sigma^3_{} 
+\frac{1}{4}g_3\sigma^4_{}$ for the scalar mesons
has turned out to be crucial \cite{BB.77}.

Using Green's function techniques it has been shown in
Ref.~\cite{KR.91} 
how to derive the relativistic Hartree-Bogoliubov
(RHB) equations from such a Lagrangian:
\begin{equation}
\left( 
\begin{array}{cc} 
 h & \Delta \\ -\Delta^* & -h^* 
\end{array} 
\right)
\left(\begin{array}{r} U \\ V\end{array}\right)_k~=~
E_k\,\left(\begin{array}{r} U \\ V\end{array}\right)_k,
\label{RHB} 
\end{equation}
$E_k$ are quasiparticle energies and the coefficients
$U_k(r)$ and $V_k(r)$ are four-dimensional Dirac spinors.
$h$ is the usual Dirac hamiltonian
\begin{equation}
h~=~\mbox{\boldmath $\alpha p$}~+~g_\omega\omega~+~ \beta(M+g_\sigma
\sigma)~-~\lambda
\label{h-field}
\end{equation}
containing the chemical potential $\lambda$ adjusted to the
proper particle number and the meson fields $\sigma$ and
$\omega$ determined as  usual in a self-consistent way from
the Klein Gordon equations in {\it no-sea}-approximation.

The pairing potential $\Delta$ in Eq. (\ref{RHB}) is given
by
\begin{equation}
\Delta_{ab}~=~\frac{1}{2}\sum_{cd} V^{pp}_{abcd} \kappa_{cd}
\label{gap}
\end{equation}
It is obtained from the pairing tensor $\kappa=U^*V^T$ and
the one-meson exchange interaction $V^{pp}_{abcd}$ in the
$pp$-channel. More details are given in Ref.~\cite{KR.91}.
There it has been found, that these forces are not able to
reproduce even in a semi-quantitative way proper pairing in
the realistic nuclear many-body problem. As in Ref.
\cite{MR.96} we therefore replace $V^{pp}_{abcd}$ in Eq.
(\ref{gap}) by the density dependent two-body force of zero
range:
\begin{equation} 
V(\mbox{\boldmath $r$}_1,\mbox{\boldmath $r$}_2) 
~=~V_0 \delta(\mbox{\boldmath $r$}_1-\mbox{\boldmath$r$}_2) 
\frac{1}{4}\left[1-\mbox{\boldmath $\sigma$}_1
\mbox{\boldmath $\sigma$}_2\right] \left(1 -
\frac{\rho(r)}{\rho_0}\right),
\label{vpp}
\end{equation}
where $r=|\mbox{\boldmath $r$}_1+\mbox{\boldmath$r$}_2|/2$.

The RHB equations (\ref{RHB}) for zero range pairing forces
are a set of four coupled differential equations for the HB
Dirac spinors $U(r)$ and $V(r)$. They are solved for the
parameter set NLSH \cite{SNR.93}, which is widely used for
the description of neutron rich nuclei, by the shooting
method in a self-consistent way. The details will be
published elsewhere. With a step size of $0.1$ fm and using
proper boundary conditions the above equations are solved
in a spherical box of radius $R \ge 15 $ fm.  The results
do not depend on the box size for $R \ge 20$ fm.  For a
radius $R=30$ fm we found the same results within an
accuracy of 0.1 \% for the binding energies and matter
radius.  Since we use a pairing force of zero range
(\ref{vpp}) we have to limit the number of continuum levels
by a cut-off energy.  For each spin-parity channel 20
radial wavefunctions are taken into account, which
corresponds for $R=20$ fm roughly to a cut-off energy of
120 MeV. For fixed cut-off energy and for fixed box radius
$R$ the strength $V_0$ of the pairing force (\ref{vpp}) for
the neutrons is determined by a calculation in the nucleus
$^{116}$Zr adjusting the corresponding pairing energy
$-\frac{1}{2}\mbox{Tr}\Delta\kappa$ to that of a
RHB-calculation using the finite range part of the Gogny
force D1S \cite{BGG.84}. For $\rho_0$ we use the nuclear
matter density 0.152 fm$^{-3}$. In order not to miss any
bound state the cut-off energy has to be larger than the
depth of the potential. But as long as this is the case and 
as long as the interaction strength is properly renormalized,
the results of this investigation stay practically
unchanged.  And in such a way, we get almost the same
result from the the pairing force of zero range and the
finite range Gogny force.

In the upper panel of Fig.~1 we show the $rms$ radii of
the protons and neutrons for the Zirconium
isotopes with mass numbers $A=110$ to $A=140$, the drip
line nucleus. They are given as
\begin{equation}
\langle r^2 \rangle^{1/2}~=~
[ \frac{1}{N_\tau}\sum_{lj}\langle r \rangle^2_{lj}]^{1/2},
\end{equation}   
where $N_\tau=Z,N$ for protons and neutrons and 
\begin{equation}
\langle r \rangle_{lj}~=~\left(\int r^2\rho_{lj} d^3r\right)^{1/2}
\label{rlj}
\end{equation}
are the contribution of the different blocks with the
quantum numbers $l$ and $j$. To guide the eye we also give
a dashed line with a $N^{1/3}$-dependence. It clearly shows
a kink for the neutron $rms$-radius at the magic neutron
number $N=82$.

These results can be understood more clearly by
considering the microscopic structure of the underlying wave
functions and the single particle energies in the canonical
basis \cite{RS.80}. Therefore, in the lower panel of Fig.~1 we
show the single particle levels in the canonical basis for the
isotopes with an even neutron number as a function of the mass
number. Going from $N=70$ to $N=100$ we observe a big gap
above the 1$h_{11/2}$ orbit. For neutron numbers larger
than the magic number $N=82$, the neutrons are filled to the
levels in the continuum or weakly bound states in the order
of $3p_{3/2}$, $2f_{7/2}$, $3p_{1/2}$, $2f_{5/2}$ and $1h_{9/2}$.  

The neutron chemical potential is given in this
Figure by a dashed line. It approaches rapidly the
continuum already shortly after the magic neutron number
$N=82$ and it crosses the continuum at $N=100$ for the
nucleus to $^{140}$Zr. In this region the chemical
potential is very small and almost parallel to the
continuum limit. This means that the additional neutrons are
added with a very small, nearly vanishing binding energy
at the edge of the continuum. The total binding energies $E$
for the isotopes above $^{122}$Zr are therefore almost
identical. This has been recognized already in Ref.
\cite{SLH.94} in RMF calculations using the BCS
approximation and an expansion in an oscillator basis,
which is definitely not reliable for chemical potentials so
close to the continuum limit. In the present investigation
we obtain the same result treating now the continuum and
its coupling to the bound orbits by pairing correlations
properly within full relativistic HB-theory in coordinate
space.

To understand the kink in the neutron $rms$-radii shown in
Fig.~1 we present in Fig.~2 the contributions $\langle r
\rangle_{lj}$ of the different quantum numbers to these
quantities as a function of the mass number.  It is clearly
seen that the radii for the negative parity levels close to
the continuum limit are responsible for the rapid increase
of the neutron radius.  Neutrons above  the closed neutron
core $N=82$ are filled into  these orbits. As more and more
neutron are added, $3p_{3/2}$ and $2_{f7/2}$ ( after $N>88$ ),
$3p_{1/2}$ ( after $N>92$ ) respectively become weakly bound,
then the contribution of further continuum $2f_{5/2}$ and
$1h_{9/2}$ become more and more important. Going from $A=122$
to $A=140$ we observe an almost constant contribution of all
the channels to the total $rms$ matter radius except a
sudden increase in the contribution of the $3p_{3/2}$,
$2f_{7/2}$, $3p_{1/2}$ and $2_{f5/2}$ channels.  This means that the
giant halo in $^{124-140}$Zr are formed by the occupation
of all these levels in the respective nucleus.

In the upper panel of Fig.~3 we show the corresponding
density distribution for neutrons and protons in the
nucleus $^{134}$Zr. By dashed lines we show calculations
for different values of the box size $R=$15, 20, 25, and
30 fm. It is clear that we need very large box sizes to
describe the halo properly. For $R=$30 fm the neutron density
is reliably reproduced only up to $r=25$ fm, where it has
decreased to 10$^{-6}$ fm$^{-3}$. The full line in the
upper panel of Fig.~3 is an asymptotic extension to
infinite box size. On the other hand, the density
distribution inside the nucleus is reproduced properly even
for small values of the box size.

In the lower panel of Fig.~3 we show the relative
contributions $\rho_{nlj}$ of the different orbits
characterized by the quantum numbers $nlj$ with respect to
the total neutron density $\rho_n$. For comparison we also
show the total neutron density in the shaded area in
arbitrary units. As we see the halo is formed essentially
by contributions from three orbits $3p_{3/2}$, $3p_{1/2}$, and
$2f_{7/2}$. The most inner part of the halo ($7\le r\le 9$ fm) the 
$2f_{7/2}$ orbit plays the dominant role. As can be seen in the
lower part of Fig.~1, this orbit is slightly below the
chemical potential and to the continuum limit in this nucleus. 
Further outside ($10\le r\le 15$ fm) its relative contribution is
strongly reduced because of the larger centrifugal barrier
felt by the $l=3$ orbit. In this region the orbit $3p_{3/2}$,
which has nearly the same position as the $2f_{7/2}$ orbit,
takes over. Because of the smaller orbital angular
momentum, this orbit feels a reduced centrifugal barrier.
For even larger distances from the center ($r \ge 15$ fm) its
relative contribution is somewhat reduced and the $3p_{1/2}$
orbit gains importance. The $3p_{3/2}$ and the $3p_{1/2}$ levels
feel the same centrifugal barrier, but the latter is situated 
directly at the continuum limit and therefore it is more loosely 
bound than the other two orbits.

In Fig.~4 we show for all the Zr-isotopes between
$A=108$ and $A=140$ the occupation probabilities in the
canonical basis of all the neutron levels near the Fermi
surface, i.e. in the interval $-20 \le E \le 10 $ MeV.
The chemical potential is indicated by a vertical line. 
For the mass numbers $A<122$ the chemical potential lies 
several MeV below the continuum limit ($E=0$) and there is 
only very little occupation in the continuum ($E>0$). The 
nucleus $^{122}$Zr has a magic neutron number and no pairing. 
As the neutron number goes beyond this closed core, the 
occupation of the continuum becomes more and more important. 
Adding up the occupation probabilities $v^2$ for the levels 
with $E>0$ we find in the continuum 2 particles for 
$N=84$, 4 for $N=86$, 6 $N=88$, roughly 3 for $N=90$, 
roughly 4 for $N=92$, roughly 3 for $N=94$, roughly 4 for 
$N=96$, roughly 5 for $N=98$, and roughly 6 for $N=100$
(where the neutron drip line is reached).

Summarizing our investigations,  we predict neutron halos
in the Zr nuclei close to the neutron drip line, which
are composed not only by one or two neutrons, as is the
case in the halos investigated so far in light $p$-shell
nuclei, but which contain up to 6 neutrons. This is a
new phenomenon, which has not been observed experimentally
so far. It would allow the study of collective phenomena in
neutron matter of low density. This prediction is based on
relativistic Hartree-Bogoliubov theory in the continuum.
It combines the advantages of a proper description of
the spin orbit term with those of full Hartree-Bogoliubov
theory in the continuum, which allows in the canonical
basis the scattering of Cooper pairs to low lying resonances
in the continuum. A density dependent force of zero range has
been used in the pairing channel. It contains no free
parameter, because its strength is adjusted for the isotope
$^{116}$Zr to a similar calculation with Gogny's force D1S
in the pairing channel. The halos are formed by two to six
neutrons scattered as Cooper-pairs mainly to the levels
$3p_{3/2}$, $2f_{7/2}$, $3p_{1/2}$, and $2_{f5/2}$. This 
is made possible by the fact that these resonances in the 
continuum come down very close to the Fermi level in these 
nuclei and by their coupling with the loosely bound levels 
just below the continuum limit.


\leftline{\Large {\bf Figure Captions}}
\parindent = 2 true cm
\begin{description}

\item[Fig.~1] Upper part: $rms$-radii for neutrons and
protons in Zr isotopes close to the neutron drip line as a
function of the mass number $A$. Lower part: single
particle energies for neutrons in the canonical basis as a
function of the mass number. The dashed line indicates the
chemical potential.

\item[Fig.~2] Contributions $\langle r \rangle_{j\pi}$ for
the different channels with the quantum numbers $l$ and
parity as a function of the mass. The left part show the
orbits with positive parity and the right part shows the
levels with negative parity.

\item[Fig.~3] Upper part: neutron and proton density distribution 
in $^{134}$Zr. Dashed lines indicate calculations for
different values of the box size $R$. Lower part: relative
contributions of the different orbits to the full neutron
density as a function of the radius. The shaded area
indicates the total neutron density in arbitrary units.

\item[Fig.~4] The occupation probabilities in the canonical
basis for various Zr isotopes with mass number A as a
function of the single particle energy. The chemical
potential is indicated by a vertical line. For $A>122$ we
also show the number $N_h$ of neutrons in the halo.
\end{description}

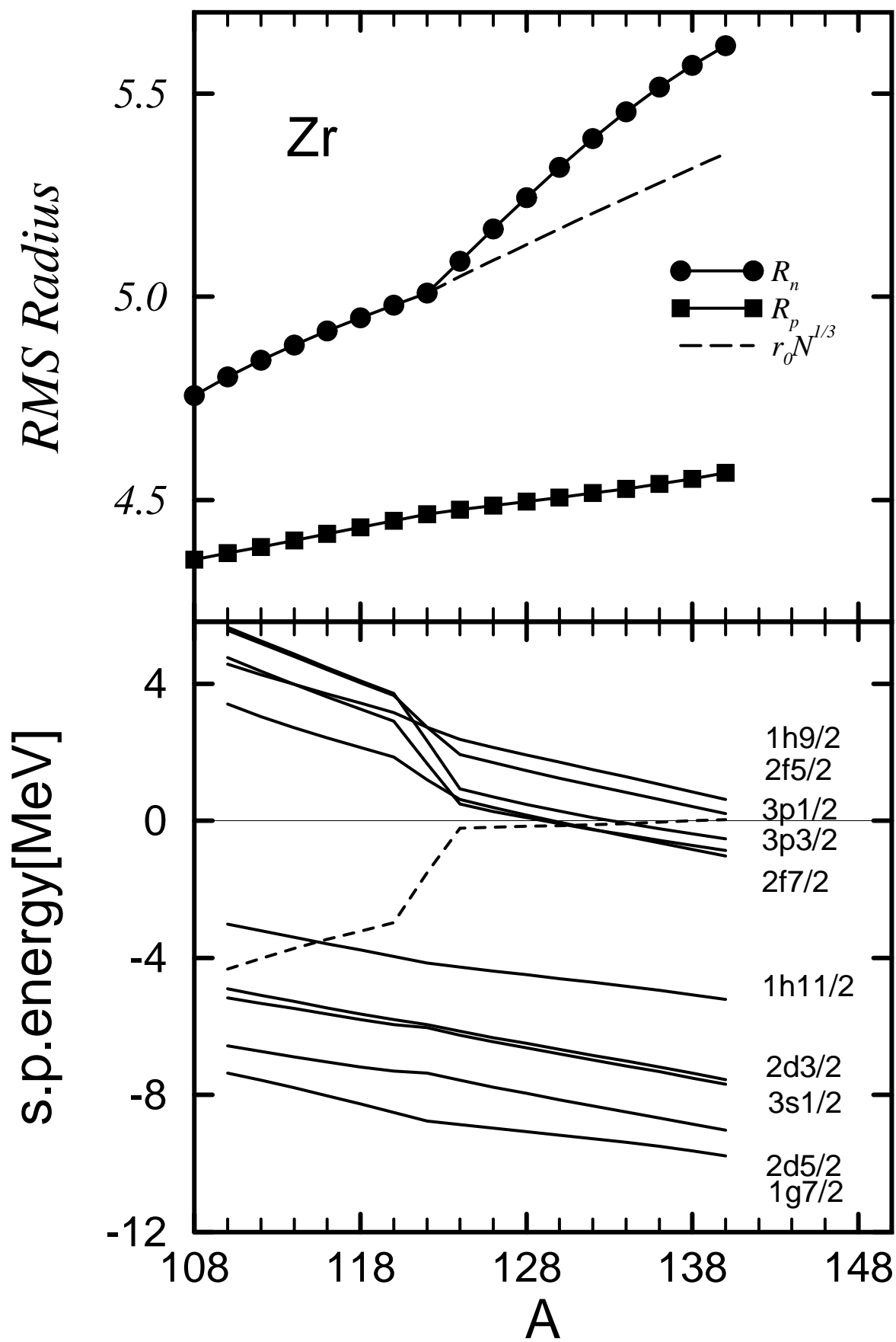
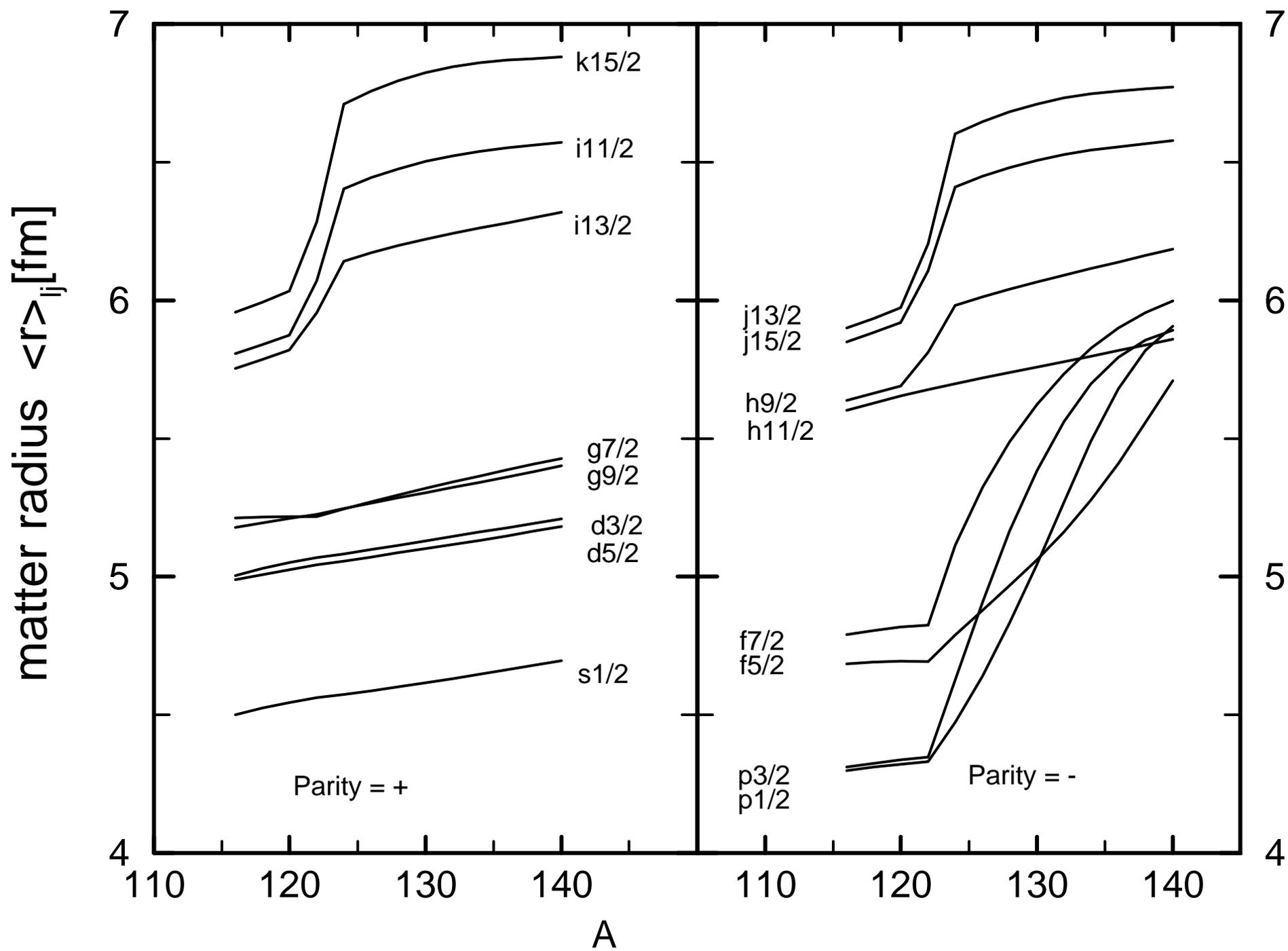
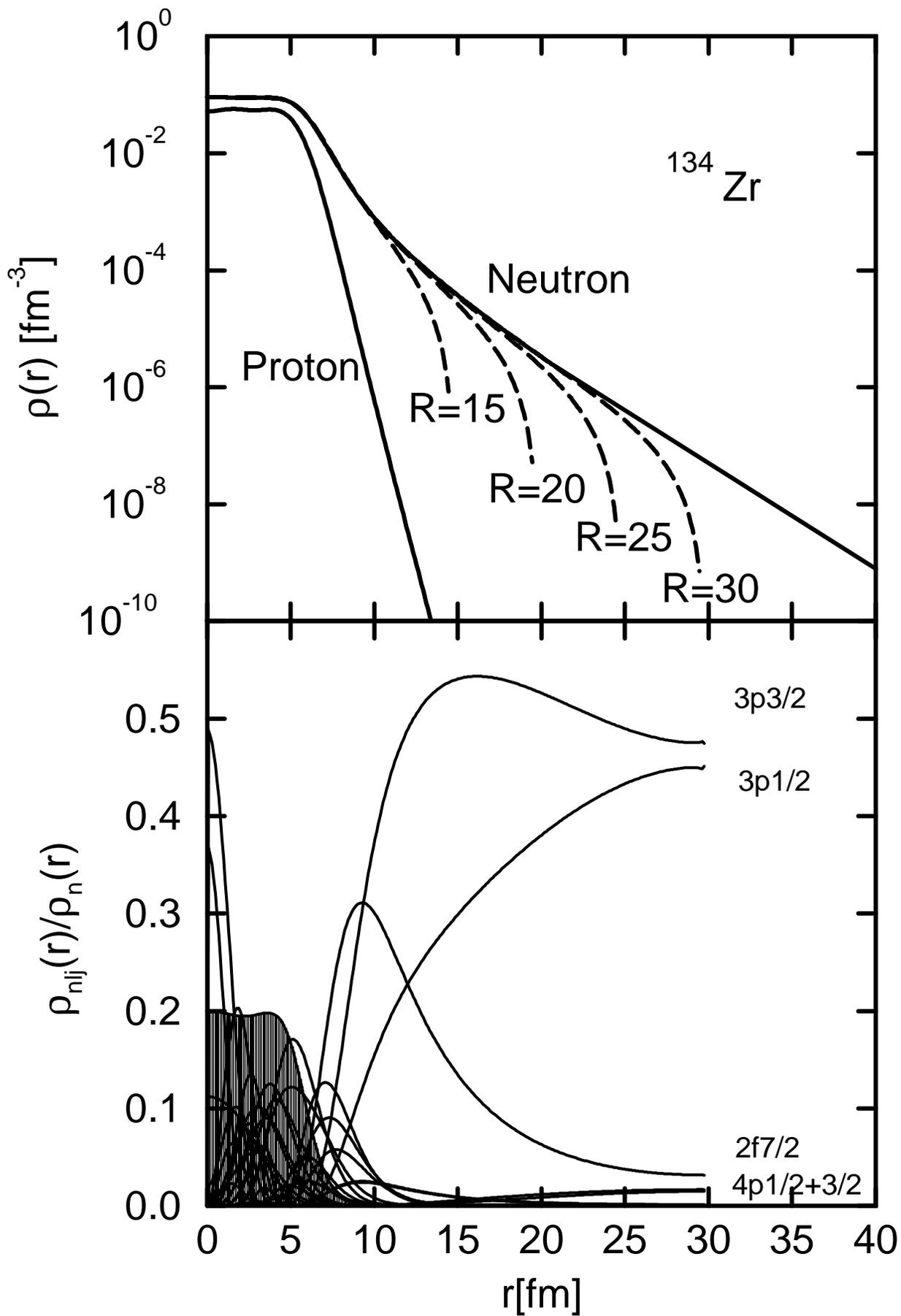
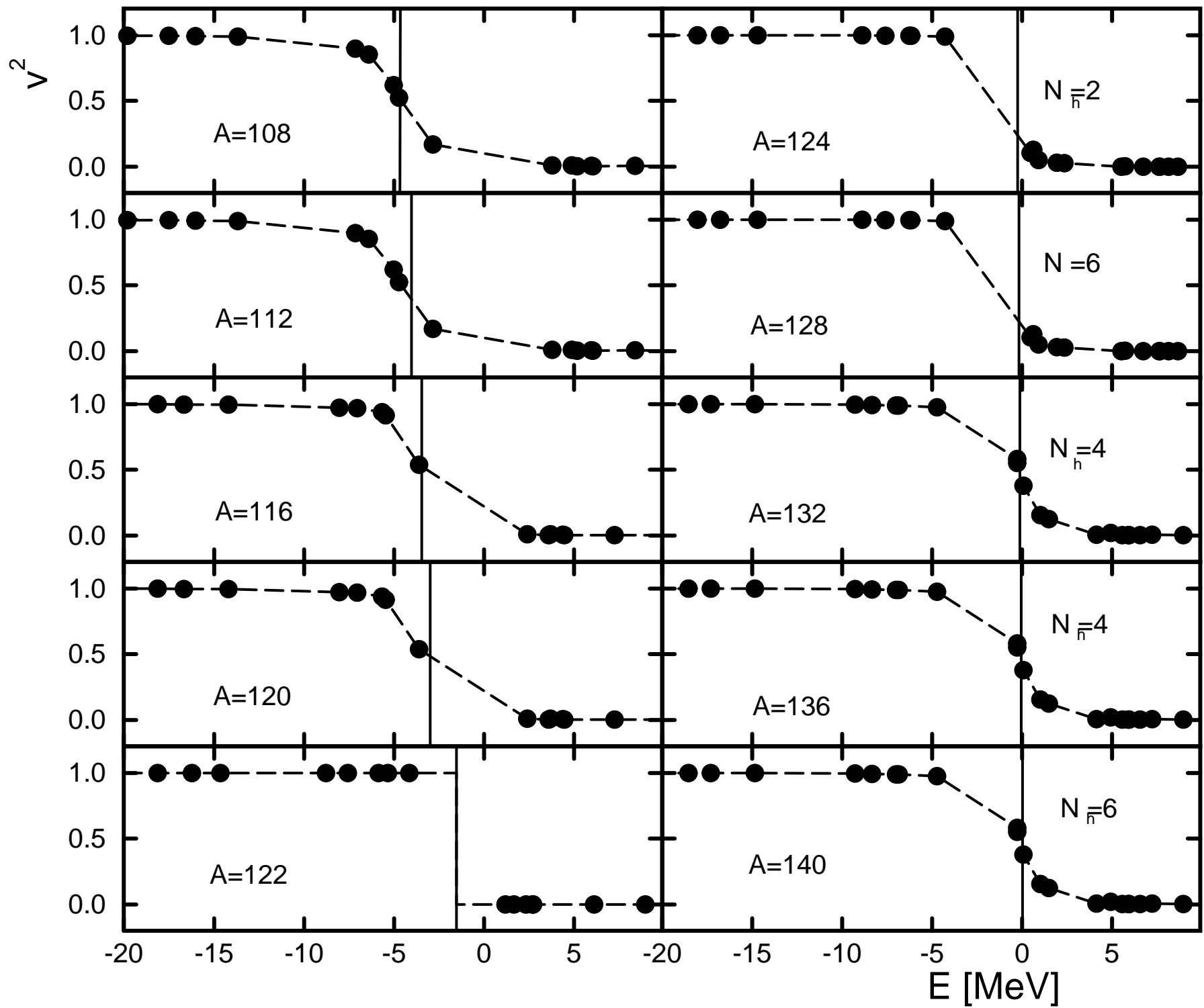
\end{document}